# Cosmic Ray Radiography of the Damaged Cores of the Fukushima Reactors.

Konstantin Borozdin,[1] Steven Greene,[1] Zarija Lukic,[2] Edward Milner,[1] Haruo Miyadera,[1] Christopher Morris,[1a] and John Perry[1]

[1]Los Alamos National Laboratory, Los Alamos, NM, USA  87544

[2]Larwence Berkeley National Laboratory, Berkeley, CA, USA 94720

**Abstract** The passage of muons through matter is dominated by the Coulomb interaction with electrons and nuclei. The interaction with the electrons leads to continuous energy loss and stopping of the muons.  The interaction with nuclei leads to angle "diffusion". Two muon-imaging methods that use flux attenuation and multiple Coulomb scattering of cosmic-ray muons are being studied as tools for diagnosing the damaged cores of the Fukushima reactors. Here we compare these two methods. We conclude that the scattering method can provide detailed information about the core. Attenuation has low contrast and little sensitivity to the core.

Shortly after the earthquake, tsunami, and core melt downs at the reactors in Fukushima Japan in March, 2011, several groups in both the United States and Japan realized that cosmic ray radiography might be able to provide information about the damaged cores. Two methods of radiography using cosmic rays have been described in the past, attenuation[1-3] and scattering.[4-6] Since deploying either of these methods to study the damaged cores of the Fukushima reactors involves a major human investment because of the high radiation fields surround the reactors, it is important to carefully evaluate the utility of the information that can be obtained from these technologies. In this paper we present a comparison of imaging using these two different techniques in a common geometry using the Monte Carlo particle transport code GEANT4.

The simulation code GEANT4[7] was used to track cosmic rays through a model of a boiling water reactor similar to Fukushima Daiichi Reactor #1.  The model of the reactor included all major structures, the reactor building, containment vessel and the pressure vessel. Calculations were performed for an intact core, a core with a 1 m diameter of material removed from the core and placed in the bottom of the pressure vessel, and with no core. A schematic view of the detector placement is shown in Figure 1. The placement of detectors outside of the reactor buildings is dictated by very high radiation levels and very limited access to the insides of the buildings.

---

[a]Electronic mail: cmorris@lanl.gov.

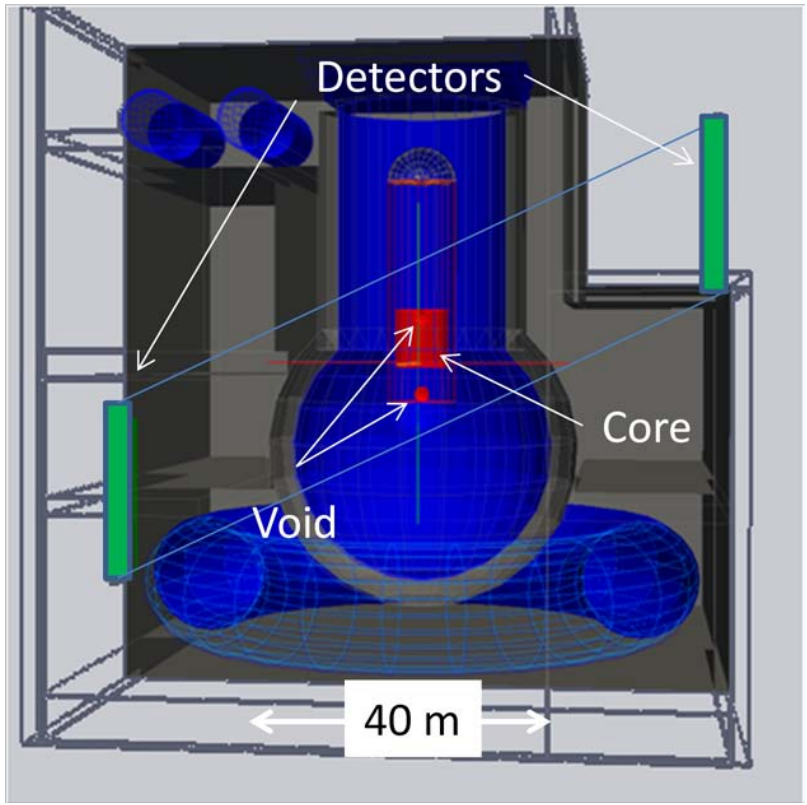

Figure 1) Cutaway view of a boiling water reactor and a schematic of the detector placement for the Monte Carlo calculation. In the case of attenuation radiography, only trajectory information from the lower detector was used. The location of the 1 m diameter void in the core and it's placement in the bottom of the pressure vessel are indicated by arrows.

Several approximations were made to simplify the calculation: Structures outside of the field encompassed by the detectors were not included (mainly the turbine buildings); the detectors were assumed to measure position and angles perfectly; there was no gamma shielding added around the detectors; the energy spectrum was assumed to be independent of Zenith angle and was taken from the 75° zenith angle measurements of Jokisch et al.[8], which corresponds to the angle of reactor core from the lower detector. A comparison of the spectra given by Jokisch et al. and by Tsuji et al.[9] shows a 50% discrepancy at low momentum and differences in the slope at higher momenta (Figure 2). This is indicative of the uncertainty in the normalization of our results.

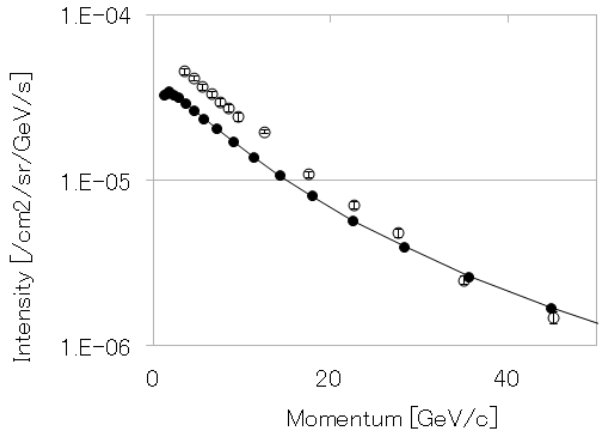

Figure 2) Cosmic ray muon energy spectrum at sea level. Solid symbols are from Jokisch[8] and the open symbols are from Tsuji[9]. Muons which penetrate the reactor lose 5-6 GeV.

The output saved from the GEANT4 runs included the input and output vectors,    and    for each incident particle. The incident flux projected to the reactor core location was used to normalize the transmission radiography (attenuation method).

The calculations are normalized to the expected 75° zenith angle flux. The muon angular distribution can be approximated by:[10]

$$\frac{dN}{d\Omega} = \frac{3}{\pi}\cos(\theta)^2 \text{ muons/min/sr/cm}^2$$

$$\Omega = \frac{\sin(\theta)hw}{l^2} \text{ sr}$$

$$N = \frac{dN}{d\Omega}\Omega\sin(\theta)hw$$

The normalization of the angular distribution gives a total muon flux of $1/\text{cm}^2/\text{min}$, when it is integrated over $2\pi$ steradians. The $\sin(\theta)$ accounts for the fact that the detectors are not normal to the line that connects their centers. The modeled detectors have h=10 m, w=5 m and l=45 m and are mounted at $\theta$=75 deg . For these conditions we expect  5.3 (2.5)×$10^5$ muons per day.

Algorithms were developed to construct images of the core using both the attenuation and multiple scattering of the cosmic rays.  The goal is to determine the sensitivity of these techniques for measuring the amount of melted fuel remained in the reactor core as well as the location of debris.

Transmission images were constructed by projecting the outgoing trajectories to a vertical plane centered in the core, and histrograming the number of events in 10x10 $\text{cm}^2$ pixels. Then the image was calculated as $-\ln(N(x,y)t_0/N_0(x,y)/t_N)$, where $N_0$ was the incident fluence and N was the transmitted fluence in exposure times of $t_0$ and $t_N$ respectively. The histogram of incident fluence was smoothed to remove an artifact introduced by the blur of the projection of the output trajectories to the plane of the core.

Plots of both the scattering images and the transmission images are shown in Figure 3 for different exposures starting at 1 hour increasing by near factors of 10 up to 6 weeks. These histograms are displayed with a linear grey scale with a lower value of zero in order make the combination of contrast and statistical fluctuations clearly visible. The times for the images are for a 50 m$^2$ detector. For a 1-m$^2$ detector these need to be increased by a factor of 50 to obtain the statistics shown at the center of the pictures. The acceptance of this geometry falls to zero at the detector edges.

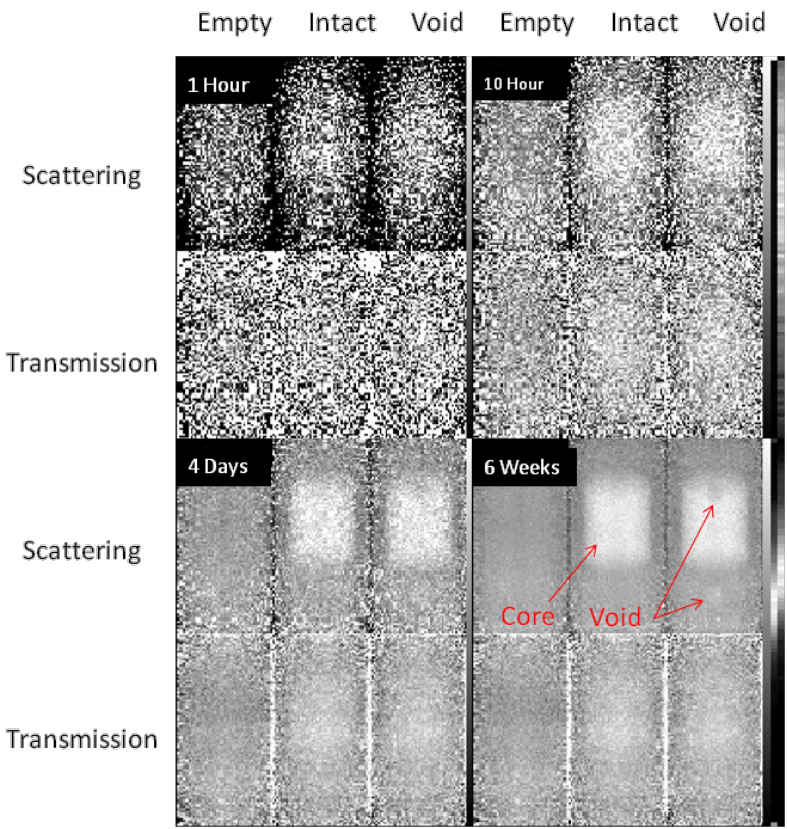

Figure 3) Reactor reconstructions at different exposure times. In scattering radiography the reactor core can be detected after about 10 hours of exposure. After four days a 1 m diameter (1%) void can be detected when compared to an intact core. After 6 weeks the void is clear and the missing material can be observed. With the attenuation method, the core can be observed when compared to an empty scene in four days. The void is undetectable even after 6 weeks of exposure.

At one hour the difference in scattering between the images with and without the core is visible and by ten hours the reactor core is visible in the scattering image. At 4 days the 1 m diameter void is visible in the core, and by 6 weeks both the void and the resulting sphere of core material the below the core are clearly visible.

The low contrast in the attenuation images is apparent when they are compared to the scattering images. The core can be detected by comparing the empty and intact images at the longer exposures, but structure in the images due to the building components shows up as strongly as the core. The void and sphere of material, clearly visible in the scattering radiograph, is not detectable in the attenuation image.

A major engineering challenge at Fukushima Daiichi is radiation shielding of deployed detectors. The site has high radiation levels on the order of one mSv/h dominantly produced by γ rays from $^{134}$Cs and $^{137}$Cs. These increase the singles counting rates and produce accidental coincidences in tracking detectors. Tests performed at the reactor site, and measurements with small scale drift tube detectors have shown that 50 cm of concrete will provide adequate shielding for operating detectors at the locations modeled here. A radiation shield of precast-concrete can enable quick installation to the site.

We have used GEANT4 to model cosmic-ray radiography of the Fukushima reactors. We have shown that 6 weeks (300 m$^2$weeks of exposure) of data provide an image with enough quality to observed 1% (a 1 m diameter sphere) of core material moved to a location below the core using scattering radiography. On the other hand, the same exposure in attenuation radiography shows far less sensitivity. This analysis shows that high quality data for radiography of the Fukushima cores from outside of the buildings can be accomplished with scattering radiography and large detectors. On site tests at Fukushima Daiichi have shown these measurements to be practical.

This work has been supported by the Laboratory Directed Research and Development program of the Los Alamos National Laboratory. We would like to thank Kanetada Nagamine for valuable discussions and Christopher Fendel for his encouragement and support.